\documentclass[12pt]{article}
\usepackage{amsmath, amssymb, mathrsfs, bbm}
\usepackage{graphicx}
\usepackage{bm}
\usepackage{amsfonts}
\usepackage{graphicx}
\usepackage{comment}
\usepackage{hyperref}

\def \rr {\raise.35ex\hbox{\small $\prime$}\kern-.17em{\mbox{\large $\imath$}}}
\def \del {\partial}
\def \dels {\partial\kern-.5em / \kern.5em}
\def \As {{A\kern-.5em / \kern.5em}}
\def \Ds {D\kern-.7em / \kern.5em}

\def \eps {\epsilon}
\def \lam {\lambda}

\def \om {\omega}

\def \one {{\bf 1}}

\def \II {I\hspace{-.1em}I\hspace{.1em}}

\def \IIB {\mbox{\II B\hspace{.2em}}}

\newcounter{multieqs}

\newcommand{\be}{\begin{equation}}
\newcommand{\ee}{\end{equation}}
\newcommand{\eq}[1]{(\ref{#1})}
\newcommand{\bit}{\begin{itemize}}  \newcommand{\eit}{\end{itemize}}

\def\bea{\begin{eqnarray}}
\def\eea{\end{eqnarray}}
\def\nn{\nonumber}

\def\one{\mbox{1 \kern-.59em {\rm l}}}

\def\a{\alpha}        
\def\b{\beta}         
\def\g{\gamma}      
\def\d{\delta}      
\def\e{\epsilon}        \def\ve{\varepsilon}

\def\k{\kappa}  
   
\def\m{\mu} \def\n{\nu}

\def\s{\sigma}

\def\cA{{\cal A}}    
  \def\cF{{\cal F}}  
 \def\cH{{\cal H}}   
  \def\cL{{\cal L}}  
\def\cM{{\cal M}}  \def\cO{{\cal O}}

\def\As {{A \hspace{-6.4pt} \slash}\;}  
  
\def\Ds {{D \hspace{-6.4pt} \slash}\;}

\def\lbb{[\hspace{-0.15em}[\hspace{0.15em}}
\def\rbb{]\hspace{-0.15em}]\hspace{0.15em}}

\def\ms{\mathfrak{s}}

\def\d{\delta}  
  
 \def\del{\partial}  
  
\def\uno{\mbox{1 \kern-.59em {\rm l}}}

\def\one{1\!\!1\,\,}

\def\RR{\mathbb{R}}

\def\bcomment#1{}  

\def\IR{\relax{\rm I\kern-.18em R}}

\begin{document}

\hfill{DCPT-10/61}

\begin{center}
\textbf{\Large
D1-brane in Constant R-R 3-form Flux and \\    
Nambu Dynamics in String Theory
}

{\large
\vskip .5in
Chong-Sun Chu$^\dagger$
\footnote{chong-sun.chu@durham.ac.uk}, 
Pei-Ming Ho$^\ddagger$
\footnote{pmho@phys.ntu.edu.tw}
}
\vskip 3mm
{\it\large
$^\dagger$ Centre for Particle Theory and Department of Mathematics, \\
Durham University, Durham, DH1 3LE, UK \\
$^\ddagger$ Department of Physics and
Center for Theoretical Sciences \\
National Taiwan University, Taipei 10617, Taiwan,
R.O.C.\\}

\vspace{60pt}

~\\
\textbf{Abstract}

\end{center}

\if
For a class of 2 dimensional field theories 
with reparametrization symmetry, 
we propose a generalized Hamiltonian formulation 
using Nambu bracket, instead of Poisson bracket,
that requires no gauge fixing.
This is a generalization of an earlier work by Takhtajan
which requires a partial gauge fixing, 
and it applies to the low energy effective theory of D1-branes 
in certain \IIB supergravity backgrounds.
\fi

We consider D1-string in a constant R-R 3-form flux background
and 
analyze its low energy limit. 
The leading order low energy theory has 
reparametrization symmetry and 
is a generalization of an earlier work by 
Takhtajan.  
We 
show that the dynamical evolution of the theory
takes a generalized Hamiltonian form in terms of a Nambu bracket.
This description is formulated in terms of reparametrization 
invariant quantities and requires no fixing of the
reparametrization symmetry.
We also show that a Nambu-Poisson $(p+2)$-bracket arises naturally 
in the 
reparametrization invariant description of the low energy theory of a 
$p$-brane in 
a constant $(p+2)$-form flux background. For example, our results
apply for a fundamental string in a constant NS-NS 3-form flux $H_3$
and 
an M2-brane in a constant 4-form flux $F_4$.

\setcounter{footnote}{0}
\newpage

\section{Introduction}

Nambu mechanics \cite{Nambu} was proposed 
as a generalization of 
the Hamiltonian formulation of classical mechanics. 
Central in its formulation 
are the replacement of the canonical phase space $(q,p)$ by a 3 dimensional
phase space consisting of a triplet 
of variables $(q^1,q^2,q^3)$ and the replacement of the 
Poisson bracket by the Nambu bracket 
\be \label{cN}
\{f_1,f_2,f_3\} = \e^{ijk} \frac{\del f_1}{\del q^i}  
\frac{\del f_2}{\del q^j} \frac{\del f_3}{\del q^k}.
\ee 
The Nambu-Hamilton equation takes the form
\be \label{NH}
\frac{df}{dt} = \{H_1, H_2,f\},
\ee
where $H_1, H_2$ are functions of $q^i$
and are called the  Hamiltonians of the system. 

In the original paper \cite{Nambu},
the properties of the canonical Nambu bracket \eq{cN}
being skew-symmetric and 
satisfying the Leibniz rule
were emphasised. Later the fundamental identity
\bea
\{g,h,\{f_1,f_2,f_3\}\} =
\{\{g,h,f_1\},f_2,f_3\} + \{ f_1, \{g,h,f_2\},f_3\}
+ \{f_1,f_2, \{g,h,f_3\}\}
\nn \\
\hfill
\forall \; g, h, f_1, f_2, f_3 \in {\cal A},
\eea
was formulated by Takhtajan
\cite{Takhtajan} and it is often accepted as a property of the Nambu bracket
(see also \cite{cz-cqN,tern1,tern2} 
for consideration otherwise).
However we note that when one generalizes 
Nambu mechanics to
a field theory with  variables $q^1(\s), q^2(\s), q^3(\s)$, where $\s$
denotes the coordinates of the base space (say $n$ dimensions),
the natural extension of \eq{cN}
\be \label{cN-field}
\{f_1,f_2,f_3\} =  \int 
d^n \s \; \e^{ijk} \frac{\del f_1}{\del q^i(\s)}
\frac{\del f_2}{\del q^j(\s)} \frac{\del f_3}{\del q^k(\s)}
\ee
does not obey the fundamental identity. The violation can be seen easily by
considering functionals $f_i$ which are nonlocal in the $q^i(\s)$'s,
e.g. an integral such as the energy.
In fact the fundamental
identity is broken in general 
for the direct
sum of two canonical Nambu brackets.
This is different from Poisson bracket where
a direct sum of canonical Poisson brackets still observes the Jacobi identity.
Apparently the canonical
Nambu bracket is more nonlinear and does not observe
a simple superposition principle.
In this paper we will consider a field theory where the bracket \eq{cN-field} 
or a direct sum of 
them naturally
determines the time evolution of the theory, much like
the role played by a Poisson bracket in the ordinary Hamiltonian formulation.
Therefore
we will not insist on the fundamental identity as a defining property
of the Nambu bracket. Instead, we will refer to
the direct sum of the canonical brackets
\eq{cN} 
as a Nambu bracket.

Nambu also showed that the Euler equation for a rotating
top can be recast into this form (\ref{NH}).  
Relation between Nambu
and Hamiltonian mechanics was clarified in the early days in \cite{early},
and also more recently \cite{cz-int}, 
where it was shown that 
Nambu mechanics of a canonical triplet can always be embedded in a Hamiltonian
system with constraint(s). If this is true in general,
Nambu mechanics
will be simply a specific form of Hamiltonian mechanics and one is compelled to
ask what are the advantages of Nambu's formulation.
One of the main results of this paper is to demonstrate that, 
at least for a class 
of theories, the description using Nambu brackets is favoured over 
the Hamiltonian 
description. 

The theories we are interested in are generalization of the 
2 dimensional field theories
\be \label{S00}
S= \int \left(\frac{1}{6}\epsilon_{ijk} q^i dq^j dq^k 
- \cH_1(q) d\cH_2(q) dt \right), \quad i,j =1,2,3
\ee
introduced by Takhtajan \cite{Takhtajan}.
An important feature of these actions is that they 
are invariant under the reparametrization 
of the spatial worldsheet coordinate $\s$
\be
t \rightarrow t' = t, \qquad
\sigma\rightarrow \sigma' = \sigma'(t, \sigma).
\label{diffeoa}
\ee
Takhtajan  showed that by partially gauge fixing this diffeomorphism
symmetry, the equation of motion of the system can be written in Nambu's form 
\eq{NH} with the use of the Nambu bracket. We refer the reader to the
appendix for a review of the theory of Takhtajan  \cite{Takhtajan}.
If one wishes, one may also
completely gauge fix the reparametrization symmetry. This allows one to 
introduce 
a Poisson structure to the theory and write the equation of motion of the 
theory in the canonical Hamilton form. 

In this paper we will consider a more general class of 
actions
\be
S = \int \left(
\frac{1}{2} C_{ij}(q) dq^i dq^j - \cH(q, q',  q'', \cdots) d \s dt
\right), \quad i,j =1,2,\cdots, D
\label{S1aa}
\ee
where $D =3n$ is the dimensions of the phase space and 
we demand that 
$\int \cH d\s dt$  (and hence the action) is 
invariant under the 
same worldsheet reparametrization (\ref{diffeoa})
\footnote{
Note that for this class of actions (\ref{S1aa}),
the invariance under the reparametrization
\be
t \rightarrow t'=t, \qquad \sigma \rightarrow \s'=\sigma'(\sigma). 
\label{residual_gauge_symm0}
\ee
implies the more general 
reparametrization symmetry (\ref{diffeoa}).
On the other hand the Hamiltonian $\int {\cal H} d\sigma$
is only invariant under (\ref{residual_gauge_symm0}).
}.
Moreover the potential $C_{ij}$ is supposed to take the canonical block 
diagonal  form
\be \label{C-block}
C_{ij} = 
\begin{cases}
f_\a \eps_{(i-3\a)(j-3\a)(k-3\a)} q^k, & \mbox{for} \quad 
i, j, k = (1+3\a, 2+3\a, 3+3\a), \\
0, & \mbox{otherwise}, 
\end{cases}
\ee
for $\a =0, \cdots, n-1$. 
We will refer to the more general action \eq{S1aa} as the 
generalized Takhtajan action. 
Since the original analysis of Takhtajan
does not apply, we will develop a new formulation 
for its dynamics.
Our formulation does not require any gauge fixing of the reparametrization 
symmetry at all and is 
based on the use of gauge invariant observables of the theory, which
is equivalent to the space of functions on the phase space. 
We will show that 
the time evolution of these gauge-invariant observables 
is naturally described in terms of a generalized 
Hamilton equation using the Nambu bracket.

Our gauge independent formulation of classical systems
provides a bridge among 
the canonical formulations in different gauges.
When a particular gauge is picked, 
the gauge fixing condition determines a specific
function $G$ on the phase space.
We will show that the Nambu bracket reduces to 
the Poisson bracket in this gauge after
plugging $G$ into one of the slots of the Nambu bracket.
Therefore 
our formulation unifies the gauge fixed descriptions of 
the theory for all gauges.

The theory \eq{S1aa} is interesting not only because it 
provides a concrete example 
illustrating the usefulness of 
the Nambu bracket.
As it turns out, the generalized Takhtajan actions 
arises naturally in string theory.
In particular, we will show that 
Nambu dynamics appears quite generically in
the low energy description of D1-string in a background with
a constant R-R 3-form field strength $F_3=dC_2$. This is another main 
result of this paper.

We should remark that the action \eq{S00} without the $\cH_1 d\cH_2$ 
term has been considered in the literature.
This describes the so called topological (open) membrane \cite{topM2,Pioline}. 
It is natural to try to embed this system in string/M theory. 
To achieve this, one needs to be able to find a suitable limit to 
decouple the bulk kinetic term.  
We will discuss 
the subtleties associated with this limit.
On the other hand, 
our low energy limit of the D1-string system does not 
suffer from these subtleties.

With a string embedding of the action \eq{S1aa}, one may ask if the
system is a well-defined quantum system on its own.  
The need of quantizing the Nambu bracket is bypassed in the gauge 
independent formulation of the theory. In this approach, 
the goal of quantization is 
to provide a quantum algebra of the gauge-invariant observables. We will 
show that 
the quantum algebra can be exactly determined
without referring to the Nambu bracket. This is yet 
another main result of this paper.


The plan of the paper is as follows. 
In section 2, we consider open M2-brane 
and open D2-brane in exact $C$-field background; 
and open D1-brane in 
a background with a constant R-R 3-form field strength $F_3$. 
We show that the generalized Takhtajan action (\ref{S1aa}),
but not the original Takhtajan action (\ref{S00}),
arises quite generically as the low energy limit of 
the D1-brane system. 
In section 3, we introduce our 
gauge-independent formulation of 
the generalized Takhtajan action. 
We show that the time evolution of the gauge 
invariant observables 
obeys a generalized Hamilton equation using Nambu bracket. We also clarify the
relation between the Nambu bracket and the Poisson bracket obtained in a 
completely gauge fixed Hamiltonian description.
In section 4, we consider the gauge independent formulation and show that
one can write down the commutator algebra of the gauge invariant observables
exactly. Higher dimensional generalization is discussed in section 5. 
Further discussions are included in section 6.

\section{Nambu Dynamics from String Theory}
\label{NambuString}

\if
In this section, we consider D1-brane in the presence of 
a constant R-R 3-form flux
and demonstrate how 
the generalized Takhtajan's action (\ref{S1aa})
arises as the leading order low energy theory of the D1-brane.
\fi

\subsection{Open M2 and D2-brane in exact $C$-field background}

In view of studying the physics of open membrane 
in the large $C$-field limit 
(or more precisely the large $H$-field limit 
for the M5-brane on which the open membrane ends),
the action \cite{topM2,Pioline}
\be
S = \int_{\cal M} C_{ijk} dX^i dX^j dX^k 
= \oint_{\del {\cal M}} C_{ijk} X^i dX^j dX^k, 
\ee
where $i, j, k = 1, 2, 3$ and 
\be
C_{ijk} = C \epsilon_{ijk}
\ee
for some constant $C$ was proposed.
In order to define the Poisson bracket, 
one has to impose a gauge fixing condition to 
break the diffeomorphism symmetry. 
Alternatively, 
one can study the algebra of
diffeomorphism-invariant observables
generated by operators of the form
\be
{\cal O}(A) = \int A_i(X) dX^i,
\ee
where $A_i(X) dX^i$ is a 1-form in the target space
\cite{Pioline}.
It was found that, 
independent of the gauge fixing condition, 
the Poisson bracket 
\be
\{ {\cal O}(A_1), {\cal O}(A_2) \} = {\cal O}(A_3)
\label{AA=A}
\ee
is isomorphic to the algebra of volume-preserving diffeomorphism. 
To illustrate the isomorphism, 
for each 1-form $A$ we define a scalar by
\be
\phi = C^{-1} \ast d A,
\ee
where $\ast$ represents the Hodge dual.
Then the algebra (\ref{AA=A}) 
can be easily seen to be equivalent to the algebra 
\be
[\d_{\phi_1}, \d_{\phi_2}] = \d_{\phi_3},
\ee
where
\be
\phi_3^i = \phi_1^j \del_j \phi_2^i - \phi_2^j \del_j \phi_1^i.
\ee
The volume-preserving diffeomorphism is generated by
\be
\d_{\phi} q^i = \phi^i.
\ee
This was first studied for arbitrary dimensions in \cite{Matsuo:2000fh}.

Turning on also the following time  components of the $C$-field,
\be \label{OHH}
C_{0ij} = - \del_i \cH_{1} \del_j \cH_{2}
\ee
and taking the temporal gauge 
\be
X^0 = \sigma^0 = t, 
\ee
the  membrane action is modified to
\be \label{M2-S}
S = \oint_{\del {\cal M}} \left(
C \eps_{ijk} X^i dX^j dX^k - \cH_1 d\cH_2 dt
\right).
\ee
This is precisely the Takhtajan action
\eq{S00}. 
Therefore understanding of the 
Takhtajan action
will be helpful for our understanding of
the physics of M2-brane. 
However the precise limit
where one can drop the mass term of the M2-brane action is subtle. We will 
comment on this later.

The system can be reduced to 10 dimensions and 
similar analysis can be performed.
Consider an open D2-brane in the presence of 
a constant axion and a 
background R-R 3-form potential
\be \label{bkgd-d2}
C^{(3)} = l_s^3 \; (C \epsilon_{ijk} dX^i dX^j dX^k - d\cH_1 d\cH_2 dX^0), 
\ee
The action is 
\bea
& & S_{D2} = S_{DBI} + S_{WZ}, \\
& &S_{DBI} = \frac{\mu}{g_s} \int d^3 \sigma \; 
\sqrt{ - \det G  }, \quad
S_{WZ} =  \mu \int \; C^{(3)} 
\eea
where $ \mu = 1/((2\pi)^2  \ell_s^3)$ is the R-R charge density 
for the D2-brane. 
Here $C$ is a  constant and 
$\cH_1$, $\cH_2$ are arbitrary functions of $X^i$, 
so that the flat spacetime background is consistent.
The indices $i, j, k$ go from $1$ to $3$.
Note that we have set the
worldvolume gauge field $A$ zero above. This is allowed by the equation of motion.

Suppose there is a limit where one can neglect the DBI term, then 
in the temporal gauge,
the D2-brane action  becomes (up to an overall constant factor) 
\be \label{d2-S}
S= \oint \left(
C \epsilon_{ijk} X^i dX^j dX^k -
\cH_1(X) d \cH_2(X) dt.
\right)
\ee
We obtain again the  Takhtajan action. 

Now we comment on the desired limit of dropping the
the bulk mass term and leaving the boundary 
term. Naively this can be achieved 
by scaling  the target space metric to zero while keeping the background 
$C$-field fixed. However 
this procedure is subtle.
Unlike open string, the  energy spectrum of  open M2-brane and 
open D2-brane system has no mass gap. Without the energy gap 
to prevent bulk excitations, it is therefore possible for
an infinitesimal deviation of the boundary excitations to
evolve into another solution with significant excitations
in the bulk. As a result
the boundary modes
will not be a good physical description of the system and we cannot trust 
the actions \eq{M2-S}, \eq{d2-S}.

\subsection{Low energy action of D1-brane in 
a constant R-R 3-form flux background}

To avoid the above problem of energy spectrum 
of the open M2-brane or open D2-brane systems, 
we consider a D1-brane system. We will now show that
the generalized Takhtajan action can  be obtained quite generically 
as the low energy effective action of a closed D1-brane in a 
background of constant RR 3-form flux. 

Consider a background with the metric $G_{\m\n}(X)$,
the dilaton $\Phi(X)$, the axion $\chi(X)$ and
the R-R 2-form gauge potential with only spatial components $C_{IJ}(X)$.
Our analysis below is valid without assuming any particular form of $C_{IJ}$. 
We will use  
$\m\n$ to denote the full spacetime indices, $0, \cdots, 9$; 
and $I, J, K$ etc to denote the spatial 
indices, $1, \cdots, 9$.
The Lagrangian density for a D1-brane is
\be
{\cal L}_{D1} = {\cal L}_{DBI} + {\cal L}_{WZ}.
\ee
Taking  a static gauge, the DBI Lagrangian for a D1-brane is
\bea
{\cal L}_{DBI} &=& - \frac{e^{-\Phi(X)}}{2\pi \a'} \sqrt{-\det(g+b+F)} \nn \\
&=& - \frac{e^{-\Phi(X)}}{2\pi \a'} 
\sqrt{(-G_{00}-\dot{\vec{X}}^2)(\vec{X}')^2 + (\dot{\vec{X}}\cdot{\vec{X}'})^2 
- (b_{01} + F_{01})^2},
\eea
where 
we have denoted
\be
\dot{\vec{X}}^2 \equiv G_{IJ}\dot{X}^I\dot{X}^J, \qquad
(\vec{X}')^2 \equiv G_{IJ}X^I{}' X^J{}', \qquad
\dot{\vec{X}}\cdot\vec{X}' \equiv G_{IJ}\dot{X}^I X^J{}'. 
\ee
In the above  we have assumed  that 
\be \label{cond-G0i}
G_{0I} = 0.
\ee
The Wess-Zumino term for the D1-brane is 
\be
{\cal L}_{WZ} = 
\frac{1}{2\pi \alpha'}
\big(C_{IJ}(X) \eps^{\a\b} \del_{\a} X^I \del_{\b} X^J
+ \chi(X) (b_{01}+F_{01})
\big) .
\label{LWZ}
\ee

The equations of motion for $A_0$ and $A_1$ are
\be
\frac{\del}{\del \s^\a}
\left[- e^{-\Phi(X)} 
\frac{b_{01}+F_{01}}
{\sqrt{-\det(g+b+F)}}
+ \chi(X) \right] = 0.
\label{EOM_A0} 
\ee
This implies that the term in the bracket $[\cdot]$ must be a constant $c$, 
and so
\be \label{bFg}
(b_{01}+F_{01})^2 = -\det g \frac{ (\chi-c)^2}{(\chi-c)^2 + (e^{-\Phi})^2}.
\ee
The $U(1)$ field $F_{01}$ is completely determined by other fields
because there is no propagating degrees of freedom
for a massless vector field in 2 dimensions.
For finite energy configurations, 
$b_{01}+F_{01}$ approaches to 0 at infinities.
From the expression above, it should then be obvious that 
one should interpret $c$ as the value of $\chi$ at the infinity of 
the 1 dimensional space.

Substituting 
\eq{bFg} back to the Lagrangian,  we obtain
\be
\cL_{D1} = -\frac{1}{2\pi \a'}K(X)
\sqrt{- \det g} + 
\frac{1}{2\pi \a'} \tilde{C}_{IJ}(X) \eps^{\a\b} \del_{\a} X^I \del_{\b} X^J,
\label{new-LDBI}
\ee
where we have defined
\bea
\tilde{C}_{IJ} &:=& C_{IJ} + cB_{IJ}, \\
K(X) &:=& 
\sqrt{(e^{-\Phi})^2+ (\chi-c)^2 }
\eea
and the negative root of \eq{bFg} is considered.
%

In the low energy approximation, 
we expand the action according to the number of
time derivatives. Up to first order in time derivatives, the action
\eq{new-LDBI} reads
\be  \label{SD1-low-energy}
S_{D1} \simeq  \frac{1}{2\pi \a'} \int \left[
\tilde{C}_{IJ}(X) dX^I dX^J - {\cal H} d\sigma dt
\right], \qquad
{\cal H} = 
K(X) \sqrt{-G_{00} (\vec{X}')^2} \;.
\ee
Note that both $\tilde{C}_{IJ}$ and $\cH$ are
independent of the time derivatives $\dot{X}^I$,
so 
the low energy action \eq{SD1-low-energy},
is invariant under the reparametrization of $\sigma$ 
\eq{diffeoa}.
So far 
$\tilde{C}_{IJ}$ is arbitrary. 
To obtain the generalized Takhtajan action \eq{S1aa}, 
we need to take it to be of the block diagonal form \eq{C-block}.
It is also needed that 
for those $X$'s which do not appear in the first term of \eq{SD1-low-energy},
denoted as $X^{i'}$,
one should be able to set them to  constants. Whether this is 
allowed depends on the equation of motion for $X^{i'}$. For example,
it is consistent to do so 
if all the background fields do not depend on $X^{i'}$. In general this 
will need to be checked case by case.


\subsection{An example: D1-brane in $\RR^3\times AdS_2 \times S^5$ 
with constant $F_3$}

In this subsection, we give 
an explicit
example of a \IIB supergravity background which satisfies the conditions
stated above and write down the 
action \eq{SD1-low-energy} explicitly. The background of interest is 
given by turning on a constant R-R 3-form flux  in the $AdS_5$ factor of the 
standard  $AdS_5 \times S^5$ background. 
As we will show below, by choosing the 
magnitudes of the R-R potentials $C_2$ and $C_4$ appropriately, 
we can determine 
the background exactly,  with the spacetime 
metric 
deformed to $\RR^3 \times AdS_2 \times S^5$. 

To show this, let us start with an ansatz with $B=0$
and with nontrivial R-R potentials $C_2$ and $C_4$. 
In the string frame, the nontrivial equations of motion are 
\bea
&& \del_\mu ( \sqrt{-G} G^{\mu \nu} \del_ \nu e^{-\Phi} ) = 0,\label{eom1}\\
&& \del_\mu ( \sqrt{-G} G^{\mu \nu} \del_ \nu \chi ) = 0,\label{eom2}\\
&& \del_\mu ( \sqrt{-G} G^{\mu \nu} F_{ \nu \g \d} ) = 0,\label{eom4}\\ 
&& R_{\mu \nu} -\frac{1}{2} R G_{\mu \nu} = 
\frac{e^{2\Phi}}{2} \left( \frac{1}{2} F_\mu{}^{\g \d} F_{ \nu\g\d} - 
G_{\mu \nu} \frac{1}{2} |F_3|^2 \right) \\
&&\qquad \qquad\qquad\qquad   +
\frac{e^{2\Phi}}{4} \left( \frac{1}{4!} F_\mu{}^{\g\d\eta\k} F_{ \nu \g\d\eta\k} - 
G_{\mu \nu} \frac{1}{2} |F_5|^2 \right) \nn \\
&&\qquad \qquad\qquad\qquad   
-4\left(\del_\mu\Phi\del_ \nu\Phi - \frac{G_{\mu \nu}}{2}(\del\Phi)^2\right)
+\frac{e^{2\Phi}}{2}\left(\del_\mu\chi\del_ \nu\chi - \frac{G_{\mu \nu}}{2}
(\del\chi)^2\right). 
\label{eom5}
\nn \\
&& F_5 = \ast F_5. \label{eom6}
\eea
Here we follow the notation and convention  of \cite{pol}. For example,
\be
|F_p|^2 = \frac{1}{p!} G^{\mu_1  \nu_1} \cdots G^{\mu_p  \nu_p} 
F_{\mu_1 \cdots \mu_p} F_{ \nu_1 \cdots  \nu_p} 
\ee
for the norm of a $p$-form.
To solve these equations,
we will take the ansatz
\be
e^{-\Phi} = \chi /(2 \sqrt{2})
\ee
so that (\ref{eom1}) implies (\ref{eom2}),
and the last two terms in (\ref{eom5}) cancel.

The self-duality equation \eq{eom6} can be solved as in the standard 
$AdS_5\times S^5$  background by considering a spacetime of the form 
$\cM_{10} = \cM_5 \times \cM_5'$ and
taking
\be
F_5 = \begin{cases} 
c \ve_5 & \mbox{on $\cM_5$}, \\
c \ve'_5 & \mbox{on $\cM_5'$}, \\
0 & \mbox{otherwise},
\end{cases} 
\ee
where $\ve_5$ and $\ve_5'$ are the volume forms on $\cM_5$ and $\cM_5'$, 
respectively. As a result, we have
 \be \label{F5F5}
\frac{1}{4!} F_\m{}^{\g\d\eta\k} F_{ \nu \g\d\eta\k} - 
G_{\m \nu} \frac{1}{2} |F_5|^2 =
\begin{cases}
-c^2 G_{\m \nu}, & \m, \nu =1,\cdots, 5 \\
c^2 G_{\m \nu}, & \m, \nu = 6, \cdots, 10. 
\end{cases}
\ee
Here $X^{1,\cdots, 5}$ (resp. $X^{6,\cdots, 10}$) 
denote the local coordinates of $\cM_5$ (resp. $\cM_5'$). 
We have assumed that $\cM_5$ is Lorentizan and hence the sign in \eq{F5F5}. 

As we will see in the next section, 
the dynamics of the system 
\eq{SD1-low-energy} is determined by a Nambu bracket, 
whose nontriviality requires 
the field strength $F_3=dC_2$ to be nontrivial. The simplest form 
of flux one can consider is
\be \label{C-ans}
C_2 = f \e_{ijk} X^i dX^j dX^k, 
\qquad i,j,k =1,2,3,
\qquad
\mbox{
where $f$ is a constant}.
\ee
This gives the field strength
\be \label{F-ans}
F_3 =
\begin{cases}
f \e_{ijk}, & i,j,k =1,2,3, \\
0, & {\rm otherwise}.
\end{cases}
\ee
The field strength $F_3$ constitutes a 
nontrivial source to the Einstein equation.
In order to have an exactly solvable background, let us consider an ansatz 
of the metric with
\be \label{flatG}
G_{ij} = \d_{ij}, \quad i,j =1,\cdots, 3.
\ee
It follows immediately that contribution of the flux \eq{F-ans} 
to the Einstein equation is a cosmological constant term:
\be \label{F3F3}
 \frac{1}{2} F_\m{}^{\g\d} F_{\n\g\d} 
- G_{\m\n}
\frac{1}{2} |F_3|^2
= \begin{cases} \frac{f^2}{2} G_{\m\n} , & \m,\n = 1,2,3,\\
 -\frac{f^2}{2}G_{\m\n} , & \mbox{otherwise}.
\end{cases}
\ee
As a result, the ansatz \eq{flatG} of a flat metric is consistent if
the contribution to $R_{\m\n}$ in \eq{F3F3} and \eq{F5F5} cancel for 
the $1,2,3$ directions. 
This requires
\be
f^2 = \frac{2}{3}c^2.
\ee
And we have
\be
R_{\m\n} = \begin{cases}
0, & \m,\n = 1,2,3, \\
- e^{2 \Phi} \frac{f^2}{2} G_{\m\n}, &\m,\n =4,5, \\
e^{2 \Phi} \frac{f^2}{4} G_{\m\n}, & \m,\n =6, \cdots, 10.
\end{cases}^3 
\ee
This has $\cM_{10} =\RR^3 \times AdS_2 \times S^5$ as  solution 
\be
ds^2 = \sum_{i=1}^3 (dX^i)^2 + R^2 (\frac{-dt^2 +dU^2}{U^2})
+ ds^2_{S^5},
\ee
where $R^2 = 2e^{-2\Phi}/f^2$ and the radius of curvature $R'$ of $S^5$ is 
determined by
$R'{}^2 = 80 e^{-2\Phi}/f^2$. 
The dilaton and axion can then be solved from (\ref{eom1}) by
\be
e^{-\Phi} = \chi /(2 \sqrt{2}) = a U
\ee
for a constant $a$.

Now a  D1-brane  placed at a constant point 
on $S^5$ and at a constant value of $U=U_0$ is consistent with the equation
of motion of the D1-brane action. 
In this case, $\chi$ is constant over the 
D1-brane and so $c$ in the
equation \eq{bFg} has to take the value $c= 2 \sqrt{2} a U_0$.  
The low energy effective
action \eq{SD1-low-energy} reads
\be \label{SD1-eg}
S= \frac{1}{2\pi \a'} \int \left[
f \e_{ijk} X^i dX^j dX^k - 
aR\sqrt{(X^i{}')^2} dt d\s
\right],
\ee
which is precisely of the form (\ref{S1aa}). Note that the $U_0$ dependence
cancelled exactly and does not appear at all in \eq{SD1-eg}. This is expected 
since otherwise there will be a nontrivial potential term depending on $U_0$ 
and the D1-brane
will not be able to sit at a constant value $U=U_0$ as allowed by the equation
of motion of the D1-brane.

\section{A New Formulation 
Without Gauge Fixing}
\label{NewFormulation}

Our results obtained above suggests us to consider field theory of the form
of \eq{S1aa}. These actions differ from 
that \eq{S00} 
of Takhtajan \cite{Takhtajan}
in that $\cH$ 
in our actions are allowed to depend on derivatives of $q$'s also. 
It was shown in \cite{Takhtajan} that by using the symmetry \eq{diffeoa} 
of the Takhtajan action, 
the equation of motion of the fundamental field $q^i$ takes the form of 
the Nambu-Hamilton equation \eq{NH}.  For 
completeness, a  review of the analysis of Takhtajan is included in the 
appendix. Our action \eq{S1aa} has 
the same reparametrization symmetry
\eq{diffeoa} 
but the analysis of Takhtajan does not apply. 
In this section 
we will propose a new formulation for the action (\ref{S1aa}).
Our formulation makes use of gauge invariant quantities and 
does not require any gauge fixing of the symmetry 
\eq{diffeoa}.
We will show that the Nambu bracket appears naturally  in 
the equation of motion of the gauge invariant observables.


\subsection{Nambu bracket and the generalized Hamilton
equation}


For simplicity, 
the action we will consider in this section is 
\be
S = \int \left(
\frac{1}{6} \epsilon_{ijk} q^i dq^j dq^k - \cH dt d\sigma
\right),
\label{Sxyz}
\ee
where $\cH=\cH(q, q', q'', \cdots)$ is the Hamiltonian density. 
It is slightly more general than (\ref{S00}) 
in the sense that 
$\cH$ is not restricted to be of the form $\cH_1^{(a)} \cH_2^{(a)}{}'$ 
for $\cH_1^{(a)}, \cH_2^{(a)}$ being functions of $q^i$ 
(but not $q^i{}'$, etc.),
although we still demand $\cH d\sigma$ to be invariant 
under the diffeomorphism symmetry (\ref{diffeoa}).
As an example, $\cH = f(q) \sqrt{q'{}^2}$ 
is not allowed for (\ref{S00}) but allowed here.
Generalizations to the cases with more than 3 $q$'s should be straightforward.

The equations of motion is given by
\be
\{ q^i, q^j \}_{wv} \equiv \dot{q}^i q^j{}' - q^i{}' \dot{q}^j
= \epsilon^{ijk} \frac{\d H}{\d q^k},
\ee
where $H = \int d\sigma \, {\cal H}$.
Here the bracket $\{\ast, \ast\}_{wv}$ is not the Poisson bracket 
for the Hamiltonian formulation of the worldvolume field theory,
but the Poisson bracket on the worldvolume coordinates.
Obviously, without 
gauge fixing,
the time evolution of local quantities such as $q^i(\s)$ 
is ill-defined.
But the complete knowledge of the dynamical system 
is already encoded in the time evolution of 
all gauge-invariant observables
of the theory.

We can build up a complete set of gauge-invariant observables 
from the single-integral observables of the form
\be
{\cal O} = \int {\cal A},
\label{OA}
\ee
where ${\cal A} = {\cal A}_i(q) dq^i$ 
is a one-form in the target space,
and $\cA_i(q)$ is a scalar with respect to the transformation \eq{diffeoa}.
The path of integration is taken to be the whole $\s$-axis and $\cO$
is a function of $t$.

Notice that if we transform ${\cal A}$ by
\be
{\cal A} \rightarrow {\cal A} + d{\lam}
\label{transf-A}
\ee
with an arbitrary function $\lam(q)$ on the target space,
the observable ${\cal O}$ is invariant.
The transformation (\ref{transf-A}) resembles 
the $U(1)$ gauge transformation of a gauge potential, 
and thus we define the ``field strength''
\be
{\cal F}_{ij}(q) = \del_i {\cal A}_j(q) - \del_j {\cal A}_i(q),
\ee
which has a one-to-one correspondence with 
single-integral observables.
Equivalently, one can use the dual vector field
\be
\phi^i(q) \equiv \frac{1}{2} \epsilon^{ijk} {\cal F}_{jk}(q),
\ee
which is a divergenceless vector in the target space,
\be
\del_i \phi^i = 0,
\ee
to label an observable.

The time evolution of $\cO$ is given by
\be
\dot{\cal O} = \int {\cal F}_{ij}(q) \dot{q}^i dq^j 
= \frac{1}{2} \int {\cal F}_{ij} \{ q^i, q^j \}_{ws} d\sigma
= \frac{1}{2} \int {\cal F}_{ij} \epsilon^{ijk} 
\frac{\d {H}}{\d q^k} d\s.
\label{dotO}
\ee
This naturally leads to the introduction of Nambu bracket 
when one recalls that a coordinate-independent expression
of the observable ${\cal O}$ is
\be
{\cal O} = \int a^{(\ms)} d b^{(\ms)}, 
\label{Oab}
\ee
where $a^{(\ms)}, b^{(\ms)}$ are 0-forms of the target space.
Eq.(\ref{dotO}) 
can then be written in the form of a generalized Hamilton
equation
\be
\dot{\cal O} 
= \{ A^{(\ms)},B^{(\ms)},  H \},
\label{HJ2}
\ee
where 
$A^{(\ms)} = \int d\sigma a^{(\ms)}$,
$B^{(\ms)} = \int d\sigma b^{(\ms)}$,
and 
$\{\ast, \ast,\ast\}$ is 
the Nambu  bracket \eq{cN-field}.
Note that the bracket is generally defined using functional derivatives 
which act on $q^i(\s)$ as well as its $\s$-derivatives.

\subsection{Nambu, Poisson  and gauge fixing}
\label{NambuPoissonGauge}

In the above we have seen that Nambu bracket is a useful device 
to encode the equation of motion when 
one does not gauge-fix the worldsheet diffeomorphism. 
On the other hand, in the ordinary Hamiltonian formulation, 
the Poisson bracket is 
well-defined only after an almost complete gauge fixing.
It is natural to suspect that there may be a relation between 
the Nambu bracket and the various Poisson brackets
obtained by different gauge fixing. Our next task is to clarify
this relation.
 
We will show that, given a gauge fixing condition, 
we can always find an observable $G$ through which 
the Nambu bracket reduces to the Poisson bracket as 
\be
\{\ast, \ast, G\} = \{\ast, \ast\}.
\ee
Note that
this kind of relation for arbitrary gauge fixing conditions
is not possible for the Takhtajan's formulation, 
because it requires a partial gauge fixing which 
may or may not be compatible 
with another gauge fixing condition.



\subsubsection{Gauge fixing and Poisson bracket}
\label{Poisson}

For simplicity of writing, we will refer to $(q^1, q^2, q^3)$ as $(x, y, z)$ in
this subsection.

Consider a generic gauge fixing condition of the form
\be
\label{gf0}
\s = \Sigma(x, y, z)
\ee
such that one of the variables, say $z$, 
can be solved in terms of $x$, $y$ and $\s$
\be
\label{gf}
z = \zeta(x, y, \s).
\ee
For example, one can use $z = \sigma$ as a gauge fixing condition.
Let us now apply the canonical formulation 
to compute the Poisson bracket. 

The equation of motion obtained from varying $S$
with respect to $z$,
\be
\label{constraint}
\dot{x}y' - \dot{y}x' - \left.\frac{\d {H}}{\d z}\right|_{z=\zeta} = 0,
\ee
where $H = \int d\s\, {\cal H}$,
becomes a constraint.
The action can then be written as
\be
S = \int dt d\sigma [\zeta(x, y, \s) (\dot{x} y' - x' \dot{y}) 
- \cH], 
\ee
where $x, y$ are independent variables 
for a given gauge fixing condition. 
To derive the equations of motion and the Poisson bracket 
from this new action, 
we vary the action with respect to $x$ and $y$,
\bea
\d S &=& \int \Big[
\d x \, (\del_x \zeta \, dx - d\zeta) \, dy 
+ \d y \, dx \, (\del_y \zeta \, dy - d\zeta) 
 \nn \\
&&  - \left(\d x \frac{\d {\cal H}}{\d x} 
+ \d y \frac{\d {\cal H}}{\d y} 
+ (\d x \del_x \zeta + \d y \del_y \zeta) \frac{\d {\cal H}}{\d z}
\right) dt d\s \Big] +
\int d\left[ \zeta \, (\d x dy - dx \d y) \right] \nn \\
&=& \int dt d\s \, \left[
\d x \, E_y - \d y \, E_x \right] (\del_\s \zeta) + 
\int dt d\s \, \del_t \left[ \zeta \, (y' \d x - x' \d y) \right], 
\label{dS}
\eea
where we have assumed that the domain of $\s$ has no boundary.
Thus the equations of motion for $x, y$ are
\bea
E_x &\equiv& \dot{x} + (\del_{\s} \zeta)^{-1}
\frac{\d}{\d y}\left(\left.{H}\right|_{z=\zeta}\right)
=0,
\label{Ex} \\
E_y &\equiv& \dot{y} - (\del_{\s} \zeta)^{-1}
\frac{\d}{\d x}\left(\left.{H}\right|_{z=\zeta}\right)
=0.
\label{Ey} 
\eea

Following Fadeev and Jackiw \cite{FJ}, 
the symplectic two-form is defined by
\be
\Omega = \d \theta, 
\ee
where $\theta$ is the total derivative term in $\d S$ (\ref{dS})
\be
\theta = \int d\sigma \zeta(x, y, \s) (y' \d x - x' \d y). 
\ee
More explicitly,
\bea
\Omega 
&=& \int d\sigma [ (- \del_y \zeta y' - \del_x \zeta x') \d x \d y
- \zeta (\d x \d y' + \d x' \d y) ] \nn \\
&=& \int d\sigma 
\left[ - \left(\frac{d}{d\s}\zeta - \del_\s \zeta\right) \d x \d y 
- \zeta (\d x \d y)' \right] \nn \\
&=& \int d\s \, (\del_\s \zeta) \, \d x \, \d y.
\eea
The coefficient $\del_\s \zeta$ in the Poisson bracket 
is always nonvanishing for any reasonable gauge fixing condition,
which must break the diffeomorphism symmetry of $\sigma$.
The Poisson bracket is thus given by
\be \label{PB}
 \{A, B\} = \int d\s \; (\del_\s \zeta)^{-1} \left[ 
 \frac{\d A}{\d x(\s)}\frac{\d B}{\d y(\s)} - 
 \frac{\d A}{\d y(\s)}\frac{\d B}{\d x(\s)} \right].
 \ee
With the Poisson bracket, the equations of motion  (\ref{Ex}), (\ref{Ey})
can be written in the Hamilton form
\be \label{ordH}
\dot{f} = \{\left. H\right|_{z=\zeta}, f\}.
\ee

\subsubsection{Reduction of Nambu to Poisson}

For the system with the action (\ref{Sxyz}), 
we learned in Sec. \ref{Poisson} that 
the Poisson bracket for the gauge fixing condition (\ref{gf}) 
is given by (\ref{PB})
\be
\{x(\s), y(\s') \} = (\del_\s \zeta)^{-1} \d(\s-\s').
\ee
On the other hand,
the Nambu bracket is given by 
\be
\{ x(\s), y(\s'), z(\s'')\}= \d(\s-\s')\d(\s'-\s'').
\ee
Comparing these two brackets, 
we find that they are related by 
\be
\label{N2P1}
\left. \{A, B, G \}\right|_{z=\zeta} 
= \{\left.A\right|_{z=\zeta}, \left.B\right|_{z=\zeta}\}
\ee
for $G$ satisfying
\be
\label{Csz}
\frac{\d G}{\d z(\s)} = (\del_\s \zeta)^{-1}(\s). 
\ee

Because $\zeta$ can be viewed as the inverse function 
of $\Sigma$ for given $x, y$,
\be
(\del_\s \zeta)^{-1} = \del_z \Sigma,
\ee
we can solve (\ref{Csz}) by 
\be
G = \int d\s \, \Sigma(x, y, z).
\label{C}
\ee

While the Nambu bracket reduces to the Poisson bracket, 
let us check that the generalized Hamilton equation 
(\ref{HJ2}) also reduces to the ordinary Hamilton equation \eq{ordH}.
Consider a gauge-invariant observable ${\cal O}$
given by (\ref{Oab}).
Using the relation (\ref{LO}) 
that will be proved 
below,
we have
\bea
\left. \dot{\cal O} \right|_{z=\zeta} 
&=& \left. \{ A^{(\a)}, B^{(\a)}, { H} \}\right|_{z=\zeta} 
= \left.[ {\cal O}, { H} ]\right|_{z=\zeta} 
= - \left.[ { H}, {\cal O} ]\right|_{z=\zeta} \nn \\
&=& - \left.\{ { H}, G, {\cal O} \}\right|_{z=\zeta} 
= \left. \{ { H}, {\cal O}, G \}\right|_{z=\zeta} 
= \{ \left.{ H}\right|_{z=\zeta}, \left.{\cal O}\right|_{z=\zeta} \}.
\eea
To see that the equality connecting the first line 
to the second line holds, 
we note that, 
due to the gauge fixing condition,
the replacement
\be
H = \int \cH d\s \rightarrow \int \cH d\Sigma
\ee
does not change the equations of motion.

\section{From the Classical to the Quantum}

Since the generalized Takhtajan action 
arises quite naturally in the low 
energy limit of D1-brane in a 
constant R-R 3-form flux background, 
it is natural to ask if this
system can be a well-defined quantum system by itself. The  quantum 
properties of the generalized Takhtajan action will be our 
next topic of discussion.
In particular we would like to understand the properties of 
the algebra of observables
in the quantum theory. 
We will show that, 
as a consequence of 
the reparametrization invariance of the theory, 
single-integral observables obey the simple 
commutation relation \eq{quantumOOO}.

\if
As a result, 
assuming the quantization is unique, 
we can use the deformation $*$-product of Kontsevich 
to write down immediately the OPE 
of reparametrization-invariant observables in the theory. 
\fi

\subsection{Classical gauge-independent algebra of observables}

In terms of the dual vector field $\phi^i(q)$, 
each single-integral observable $\cO$  
can be mapped to an operator $\hat{\cal O}$ 
acting on the 
vector space spanned by all 
gauge-invariant quantities
\be \label{cO}
\hat{\cal O} \equiv \int d\sigma \; \phi^i(q) \frac{\d}{\d q^i(\s)}.
\ee
The generalized Hamilton equation (\ref{HJ2}) 
can then be written as
\be
\dot{\cal O} = \hat{\cal O}{ H}.
\ee
The space of the operators \eq{cO} is 
equipped with a Lie algbera structure
\be \label{lie-cO}
[\hat{\cO}_{\phi_1}, \hat{\cO}_{\phi_2}] = \hat{\cO}_{\phi_3},
\ee
where $\phi_3$ is the  dual vector field
\be
\phi_3^i = \phi_1^j \del_j \phi_2^i 
- \phi_2^j \del_j \phi_1^i.
\label{phi3}
\ee 
This naturally induces on the vector space of  
single-integral observables  a Lie algebra structure defined by
\be \label{lie-O}
[\cO_{\phi_1}, \cO_{\phi_2}] = \cO_{\phi_3}.
\ee
Using the bracket, one can rewrite 
the generalized Hamilton equation as
\be
\dot{\cal O} = [ {\cal O},  H ],
\ee
as the Hamiltonian $H$ is also a single-integral observable.

As discussed in the previous section,  one can fix a gauge and obtain the 
Poisson bracket \eq{PB}. 
When restricted to gauge invariant quantities, the
Poisson bracket is the same as the Lie bracket \eq{lie-O} \cite{Pioline}.
Furthermore, since the Lie algebra  \eq{lie-O} is 
equivalent to the algebra 
of volume-preserving diffeomorphism in the target space with
\be \label{vpd}
\d q^i = \phi^i(q), \qquad
\del_i \phi^i =0,
\ee
and since a gauge-invariant observable is in general a function 
of single-integral observables, 
we can identify the classical algebra of all gauge-invariant observables 
with the universal enveloping algebra of volume-preserving diffeomorphisms.
We emphasis that, just as area preserving diffeomorphism of the 
standard phase space leave the Poisson bracket invariant, 
the volume-preserving diffeomorphism \eq{vpd}  is a
symmetry of the Nambu bracket.
However,  \eq{vpd} is generally not 
a symmetry of the theory 
(discussed in the appendix \ref{symm}).

We also note that 
the Lie bracket can be
directly related to the Nambu-bracket \eq{cN-field}. In fact,
for any two single-integral  observables 
\be
{\cal O}_{\phi_1} = \int a_1^{(\ms)} db_1^{(\ms)}, \qquad
{\cal O}_{\phi_2} = \int a_2^{(\ms)} db_2^{(\ms)},  \qquad
\ee
the Lie bracket \eq{lie-O} gives
\be
[ {\cal O}_1, {\cal O}_2 ] =  - \int  \eps_{ijk} \phi_1^j \phi_2^k
dx^i
= \left\{ \int d\s \, a_1^{(\ms)}, \int d\s \, b_1^{(\ms)}, {\cal O}_2 \right\}
\label{LO}
\ee
where $\{\ast,\ast,\ast\}$ is the Nambu bracket \eq{cN-field}.

\if
One can check that \eq{LO} indeed defines
a Lie algebraic structure
\be
[{\cal O}_{\phi_1}, {\cal O}_{\phi_2}] = 
\int D\phi_3 \; f_{\phi_1\phi_2}^{\phi_3} {\cal O}_{\phi_3}
\label{OOO}
\ee
with the structure constant given by
\be
f_{\phi_1\phi_2}^{\phi_3} =
\delta(\phi_3 - \phi_1\cdot\del \phi_2 + \phi_2\cdot\del \phi_1).
\ee
The Dirac delta function $\delta$ used here is defined so that 
\be
\int D\phi \; \delta(\phi - \phi_0) {\cal O}_{\phi} = {\cal O}_{\phi_0}.
\ee

As the Nambu bracket reduces to a Poisson bracket 
upon gauge fixing, 
the 2-bracket (\ref{LO}) captures the gauge-invariant structure 
of Poisson algebra for any gauge fixing condition $\Sigma = \sigma$.
This algebra is also isomorphic to the algebra of 
the corresponding operators $\hat{\cal O}$
\be
[{\cal O}_1, {\cal O}_2] = {\cal O}_3
\qquad \leftrightarrow \qquad
\hat{\cal O}_1\hat{\cal O}_2 - \hat{\cal O}_2\hat{\cal O}_1 
= \hat{\cal O}_3.
\ee

\fi

\subsection{Quantization of gauge-invariant observables}
\label{algebra}


When one quantizes the theory, the Poisson bracket relation \eq{lie-O}
of gauge invariant observables becomes the commutator algebra
\be 
\lbb \mathbb{O}_{\phi_{1}}, \mathbb{O}_{\phi_{2}}\rbb
\equiv
\mathbb{O}_{\phi_{1}} \mathbb{O}_{\phi_{2}}
- \mathbb{O}_{\phi_{2}} \mathbb{O}_{\phi_{1}}
= i \hbar \; \mathbb{O}_{\phi_{3}}.
\label{quantumOOO}
\ee
Here $\mathbb{O}(\phi)$ is the quantized operator corresponds to 
the classical observable ${\cal O}(\phi)$.
We used the notation $\lbb \ast, \ast \rbb$
for the commutator in order to distinguish it from 
the bracket $[\ast, \ast]$ (\ref{lie-O}) for the classical algebra. 
To obtain the relation \eq{quantumOOO} directly,
one can fix a gauge 
and apply the resulting canonical equal time commutation relation. 
In principle, due to operator singularities, 
there could appear additional terms (like Schwinger terms in current algebra) 
on the right hand side of \eq{quantumOOO}. 
Yet unless the gauge symmetry 
\eq{diffeoa} is broken by an anomaly, which 
cannot be the case here since we did 
not include any chiral fermions, 
the additional terms must be gauge invariant.
Since due to locality,
the additional terms (if there)
must be of the form of single integrals, 
the relation is only modified to the form
$
\lbb \mathbb{O}_{\phi_{1}}, \mathbb{O}_{\phi_{2}}\rbb =
 i \hbar \; \mathbb{O}_{\phi_{3}} + \sum_{\phi'} 
f_{\phi_1 \phi_2}{}^{\phi'} \mathbb{O}_{\phi'},
$
with some structure constant 
$f_{\phi_1 \phi_2}{}^{\phi'} $.
Now it is known that the VPD algebra is rigid 
\cite{rigid}, 
that is, it admits no non-trivial deformation. 
Therefore, 
any additional terms to \eq{quantumOOO} 
can always be absorbed by redefinitions of the generators,
and the commutator algebra takes the form
\eq{quantumOOO} without loss of generality.



The result \eq{quantumOOO}  
agrees in general with the canonical quantization.
We will illustrate this explicitly using an example.

For the action (\ref{Sxyz}), the Nambu bracket is
\be
\{ x(\s), y(\s'), z(\s'') \} = \d(\s-\s') \d(\s'-\s''),
\ee
regardless of what the Hamiltonian ${\cal H}$ is.
(Here $(q^1, q^2, q^3)$ is also referred to as $(x, y, z)$.)
We impose the gauge fixing condition
\be
z(\s) = \s,
\ee
and then the Nambu bracket reduces to the Poisson bracket
\be
\{ x(\s), y(\s') \} = \d(\s-\s').
\ee
The canonical quantization for this gauge is given by
\be
[ \hat{x}(\s), \hat{y}(\s') ] = i\hbar \d(\s-\s').
\label{Can-xy}
\ee

For a basis of functions $f_m(z)$ for the $z$-dependence,
or equivalently 
the $\s$-dependence,
we define 
\be
x_m \equiv \int d\s \, x(\s) f_m(\s), \qquad
y_m \equiv \int d\s \, y(\s) f_m(\s).
\ee
Assuming that $\{f_m\}$ is an orthonormal basis, 
the canonical quantization (\ref{Can-xy}) 
can be equivalently expressed as 
\be
[ \hat{x}_m, \hat{y}_n ] = i\hbar \d_{mn}.
\label{Can-xmyn}
\ee
Note that functions of $x_m, y_n$ constitute 
all observables of the theory in this gauge.

Let us show how these commutation relations are reproduced 
in our formulation.
Parallel to the
general discussion above,
in our gauge-invariant formulation, 
we define the single integral observables
\be
{\cal O}_{x_m} \equiv \int dz \, x f_m(z),
\qquad
{\cal O}_{y_m} \equiv \int dz \, y f_m(z).
\ee
The bracket defined by (\ref{LO}) gives
\be
[{\cal O}_{x_m}, {\cal O}_{y_n}]
= \int dz \, f_m(z) f_n(z).
\ee
The right hand side is a single-integral observable.
On the other hand, by definition of $f_m$,
the right hand side should equal $\d_{mn}$.
The identification
\be
\int dz \, f_m(z) f_n(z) = \d_{mn} 
\ee
is consistent with the operator algebra only if 
the left hand side and the right hand side commute 
with all other operators in the same way. 
It can be checked that indeed the left hand side 
is a central element for the bracket and so 
we can safely use this identity.
As a result, upon quantization, we have
\be
\lbb \mathbb{O}_{x_m}, \mathbb{O}_{y_n} \rbb 
= i\hbar \d_{mn}.
\ee
This is exactly the same as the canonical commutation relations
(\ref{Can-xmyn}) through the map
\be
\hat{x}_m \leftrightarrow \mathbb{O}_{x_m}, \qquad
\hat{y}_m \leftrightarrow \mathbb{O}_{y_m}.
\ee

As we have commented, 
all functions on the phase space are
functions of $x_m, y_m$ in the gauge $z = \s$.
In the above we showed explicitly how the canonical quantization 
for the gauge $z = \sigma$ is entirely embedded in our formulation
without missing any observable.



\section{Higher Dimensions}

The generalization of our analysis to higher dimensions is straightforward.
Consider the following action 
\be
S = \int d^{p+1} {\cal L} = 
\int d^{p+1}\sigma \left[
-\sqrt{-\det g} + \frac{1}{(p+1)!} C_{\mu_1\cdots\mu_{p+1}}
 \eps^{\a_1\cdots\a_d}\del_{\a_1}X^{\m_1} \cdots \del_{\a_d}X^{\m_{p+1}},
\right],
\label{pbrane_action}
\ee
where
\be
g_{\a\b} = \del_\a X^{\m} \del_\b X^{\n} G_{\m\n}
\ee
and the spacetime metric is $G_{\m\n}$. 
The action is the bosonic part of the super $p$-brane action
\cite{super}
and is a generalization 
of the action of a string and that of a membrane to higher worldvolume 
dimensions

Consider a $C$-field with only spatial
components. Assuming that the metric satisfies
\be
G_{0I} =0.
\ee
In the leading order of the
low energy limit where we ignore any time derivatives of order higher than one, 
we have
\be
\cL \simeq
-\sqrt{-G_{00} \det(G_{IJ}\del_a X^I \del_b X^J)} + \frac{1}{(p+1)!}
C_{I_1\cdots I_{p+1}}
 \eps^{\a_1\cdots\a_d}\del_{\a_1}X^{I_1} \cdots \del_{\a_d}X^{I_{p+1}},
\ee
where $a,b =1, \cdots, p$ denote the spatial indices of the worldvolume.
This leads us to consider $(p+1)$-dimensional action of the form
\be
S= \int \left[
\frac{1}{(p+1)!}
C_{I_1\cdots I_{p+1}} dX^{I_1}\cdots  dX^{I_{p+1}} - \cH \;d^p \s dt
\right],
\ee
where $\cH$ depends on $X^I, \del_a X^I$ etc 
and the action is invariant under the reparametrization
\be \label{higher-repara}
t \to t'=t, \qquad 
\s^\a \to \s'{}^a = \s'{}^a(t, \s^b)
\ee
of the worldvolume. In particular we are interested in the action
\be
S= \int \left[
\frac{1}{(p+2)!}
f \e_{I_1\cdots I_{p+2}} X^{I_1} dX^{I_2}\cdots  dX^{I_{p+2}} - \cH \;d^p \s dt
\right],
\ee
which corresponds to the case 
of a constant field strength. This is the
generalized Takhtajan action in higher dimensions.

The equation of motion is given by ($f=1$)
\be
\{ X^{I_1}, \cdots, X^{I_{p+1}}\}_{wv}
\equiv
\e^{\a_1 \cdots \a_{p+1}} \del_{\a_1} X^{I_1} \cdots \del_{\a_{p+1}} X^{I_{p+1}}
= \e^{I_1 \cdots I_{p+1} I_{p+2}}
\frac{\d H}{\d X^{I_{p+2}}},
\ee 
where $H = \int {\cal H} d^p\s$.
Without fixing the reparametrization symmetry \eq{higher-repara}, 
the time evolution of local quantities such as $X^I(\s)$ is ill-defined.
Nevertheless the time evolution of gauge invariant quantity is well-defined as 
in the $p=1$ case. Consider gauge invariant observables of the form 
\be
\cO = \int \cA,
\ee
where $\cA = \cA_{I_1 \cdots I_p}(X) dX^{I_1} \cdots d X^{I_p}$ is a $p$-form 
in the target space and 
the integration is over the spatial part of the worldvolume. It is
\be 
\dot{\cO} = 
\int \cF_{I_1 \cdots I_{p+1}} \dot{X^{I_1}} dX^{I_2}\cdots dX^{I_{p+1}}
= \frac{1}{p+1} \int \cF_{I_1 \cdots I_{p+1}}  \e^{I_1 \cdots I_{p+1} I_{p+2}}
\frac{\d H}{\d X^{I_{p+2}}},
\ee
where $\cF = d\cA$.
This can be naturally written in terms of  a Nambu-Poisson $(p+2)$-bracket:
\be
\{ f_1, \cdots, f_{p+2} \} \equiv \int d^p \s 
\e^{I_1 \cdots I_{p+2}} 
\frac{\d f_1}{\d X^{I_1}(\s)} \cdots \frac{\d f_{p+2}}{\d X^{I_{p+2}}(\s)}.
\ee
Indeed for  a coordinate independent expression of the observable $\cO$
\be
\cO = \int a_1^{(\ms)} da_2^{(\ms)} \cdots da_{p+1}^{(\ms)},
\ee
the time evolution for $\cO$ reads
\be
\dot{\cO} = \{ A_1^{(\ms)}, A_2^{(\ms)} ,\cdots, A_{p+1}^{(\ms)}, H\}
\ee
where $A_i^{(\ms)} = \int d^p \s  a_i^{(\ms)}$.

We note that the above analysis does not immediately apply to 
D$p$-branes since we have not included a worldvolume gauge
field. The only exception is the D1 case where as we have shown in section 2,
the worldvolume gauge field  can be solved in terms of the other 
degrees of freedom of the theory and hence does not appear in the 
low energy action.  
However there are many branes whose worldvolume actions take 
the form of \eq{pbrane_action}.
For example our analysis holds for 
an M2-brane
in the presence of a constant 4-form flux  or a fundamental string
in the presence of a constant NS-NS 3-form flux. Our result states that
a Nambu-Poisson 4-bracket or a Nambu bracket 
arises naturally in the low energy description of these
theories.

We also note that our result is different from 
other results \cite{alg} where a 
Nambu-Poisson $(p+1)$-bracket 
was found to be useful 
in writing the action of string or  super $p$-branes. 

\section{Discussions}


In this paper we did not discuss the  problem of the
quantization of the Nambu bracket. In fact we were able to determine 
exactly the algebra of observables \eq{quantumOOO} without the need to 
quantize the generalized Takhtajan theory explicitly. 
Any valid quantization is expected to 
reproduce the result \eq{quantumOOO}.
A quantization of the Nambu bracket in terms of 
the Zariski quantization has been proposed in \cite{dito}. 
It will be interesting
if it is possible to make connection with
the result \eq{quantumOOO} explicitly.
Generalizations of the results of quantization to theories 
using higher order Nambu-Poisson bracket will also be interesting.
We expect that the algebra of volume-preserving diffeomorphism 
(for higher dimensional volume) will appear.

It is intriguing that Nambu bracket or a Nambu-Poisson bracket of higher order 
appears naturally 
in the gauge invariant description of the low energy dynamics of
branes in string theory. 
Obviously through dimensional reduction, the 
Nambu-Poisson bracket of different orders can be connected with
each other, just as the brane theories of different worldvolume dimensions do.
For example, the Nambu-Poisson 4-bracket which appears 
in the low energy description of
M2-brane in a constant 4-form flux 
is related to the Nambu bracket   in the low energy description of a 
fundamental string in a constant NS-NS 3-form flux by dimensional reduction.
It will be interesting to clarify further the role of these brackets in the 
physics of string and M-theory. For discussions of the role of Nambu
bracket in M-theory, see for example \cite{hoppe,minic,pm}.


Recently, there is some interest in the Lie 3-algebra 
as a novel way of describing symmetries.
A Lie 3-algebra is equipped with a Lie 3-bracket, 
which, like the Nambu bracket, 
is defined as a map 
$( \cdot, \cdot, \cdot ): \cA^{\otimes^n} \to \cA$
for a linear space $\cA$,
and is skew-symmetric.
Furthermore, the Lie 3-bracket is required to satisfy
the fundamental identity
\bea
(g,h,(f_1,f_2,f_3) ) = 
( (g,h,f_1),f_2,f_3) + ( f_1, (g,h,f_2),f_3)
+ (f_1,f_2, (g,h,f_3))
\nn \\
\hfill
\forall \; g, h, f_1, f_2, f_3 \in {\cal A}.
\eea
The fundamental identity ensures that 
the bracket can be used to generate a Lie algebra 
with generators labelled by two elements in $\cA$ as
\be
L(f_1, f_2) = (f_1, f_2, \ast).
\ee
For $\cA$ being the space of functions on a manifold,
a Nambu-Poisson bracket is a Lie 3-algebra which also satisfies the Leibniz rule 
\be
(f_1 f_2, g,h) = f_1 ( f_2, g, h) + (f_1, g, h) f_2
\ee
in addition to the fundamental identity.
It is tempting to consider the Nambu-Poisson bracket 
as the Nambu bracket in a generalization of the canonical formulation.
It is also very tempting to demand that the quantum version 
of the Nambu bracket be a Lie 3-algebra.
However, 
efforts in this direction have not been fruitful.
In our formulation, 
we do not demand the fundamental identity on the Nambu bracket. 
The only purpose of the Nambu bracket 
is to deliver a generalized Hamilton equation 
that determines the time evolution of all gauge-invariant observables.
We have shown in this paper that, 
despite the absence of the fundamental identity, 
the Nambu bracket is useful for a generalized Hamiltonian formulation 
of 2 dimensional field theories with 
a reparametrization symmetry 
in the spatial coordinate $\sigma$.
Our formulation does not require a choice of gauge fixing, 
but the Nambu bracket allows us to define the 3-algebra 
as an analogue of the Poisson algebra.

It will be very interesting to 
generalize our formulation to other theories with reparametrization symmetry, 
for example,
gravitational theories.
Although the problem of UV divergence is not expected to 
be alleviated,
we hope that it may bring us new insights to 
the quantum nature of gravity.

\section*{Acknowledgment}

CSC is grateful to David Fairlie for useful discussions on Nambu bracket.
The work of CSC is supported by STFC.
The work of PMH is supported in part by
the National Science Council,
and the National Center for Theoretical Sciences, Taiwan, R.O.C.

\appendix
\section{Review and Extension of 
Takhtajan's Formulation using Nambu Bracket}
\label{Takhtajan}

The application of Nambu bracket to
a certain class of 2-dimensional theories 
with diffeomorphism symmetry was first proposed 
by Takhtajan \cite{Takhtajan}. 
In this appendix we review and give a minor extension of 
Takhtajan's formulation of
a class of 
2 dimensional field theories with 
diffeomorphism symmetry.
Our new formulation of 
an even more general class of theories 
is given in Sec. \ref{NewFormulation}.

\subsection{Nambu bracket via Takhtajan} 

The action Takhtajan considered 
is (\ref{S00})
\be \label{S0}
S= \int (x dydz - \cH_1 d\cH_2 dt),
\ee
where $\cH_1, \cH_2$ are functions of on the 3-dimensional phase 
space $\RR^3$ with coordinates $(q^1,q^2,q^3)= (x,y,z)$. 
The action \eq{S0} can be
compared with the usual action functional 
\be
S= \int (p dq - \cH dt)
\ee
for Hamiltonian mechanics in a 2 dimensional phase space. 
The generalization to 
a $2n$ dimensional phase space is straightforward by simply replacing 
the 1-form $pdq \to p_i dq_i$. 
This motivates us to consider 
the following 
more general form 
of the Takhtajan action
\be
S = \int \left(
\frac{1}{2} C_{ij}(q) dq^i dq^j - \cH_1^{(a)}(q) d\cH^{(a)}_2(q) dt
\right), 
\label{S1a}
\ee
where $i, j = 1, 2, \cdots, D$ 
and $D=3n$.
The $q$'s are functions of 
the worldsheet coordinates $(t, \s)$ 
and $\cH_1$ and $\cH_2$ are 
functions of $q^i$.
We have generalized the action to higher dimensions 
and introduced the potential $C_{ij}$. 
We have also generalized 
the target space 1-form $\cH_1 d \cH_2$ to the 1-form
\be
\om \equiv \cH^{(a)}_1 d \cH^{(a)}_2, \quad a = 1, 2, \cdots, N,
\label{om-def}
\ee
in the cohomology,
that is, it is defined up to exact 1-forms 
since the Lagrangian is defined up to total derivatives and we will
consider world-sheet without boundary in this paper. 
The action \eq{S1a} is 
invariant under 
an $O(N)$ global symmetry rotating 
the index $(a)$.
It is also invariant under 
world-sheet coordinate transformations of the form 
\be
t \rightarrow t' = t, \quad
\s \rightarrow \s' = \s'(t, \s).
\label{diffeoa1}
\ee 
This is a gauge symmetry of the theory.

The equation of motion  of the 
Takhtajan action \eq{S1a} reads
\be
 q^j{}' 
(F_{ijk} \dot{q}^k - \del_i \cH_1^{(a)} \del_j \cH_2^{(a)} 
+ \del_j \cH_1^{(a)} \del_i \cH_2^{(a)}) 
= 0,
\label{EOM1a}
\ee
where 
\be
F_{ijk} \equiv \del_i C_{jk} + \del_j C_{ki} + \del_k C_{ij}.
\ee
We will now employ the  same analysis as Takhtajan \cite{Takhtajan} and 
show that upon a partial gauge fixing of 
the gauge symmetry \eq{diffeoa1}, 
the equation of motion (\ref{EOM1a}) 
is equivalent to the following 
Nambu-Hamilton equation 
\be
\dot{q}^i(\s) - \{ H_1^{(a)},  H_2^{(a)}, q^i(\s)  \}= 0, 
\label{HJ1}
\ee
where 
\be
 H_i^{(a)} \equiv \int d\s \,  \cH_i^{(a)}(q(\s)), \qquad i = 1, 2,
\ee
and
the 3-bracket $(\ast,\ast,\ast)$ 
is defined by $F_{ijk}$ in the following way:
We will be interested in the case where
$F_{ijk}$ takes the canonical block diagonal form
\be \label{fijk}
F_{ijk} = 
\begin{cases}
f_\a \eps_{(i-3\a)(j-3\a)(k-3\a)}, & \mbox{for} \quad 
i, j, k = (1+3\a, 2+3\a, 3+3\a), \\
0, & \mbox{otherwise}.
\end{cases}
\ee
Here $f_\a$ are constants and $\a =0,\cdots, n-1$.
In this case one can introduce a 3-bracket such that 
$\{ q^i(\s), q^j(\s'), q^k(\s'') \} $ is nonvanishing only 
for $i, j, k \in (1+3\a, 2+3\a, 3+3\a)$ 
for a given value of $\a$:
\be \label{qqq}
\{ q^i(\s), q^j(\s'), q^k(\s'') \}= 
f^{-1}_\a \eps^{(i-3\a)(j-3\a)(k-3\a)} \d(\s-\s') \d(\s'-\s'').
\ee
That is,
\be \label{Nfff}
\{g_1,g_2,g_3\} =  \int d\s \; \sum_\a 
f^{-1}_\a
\sum_{\stackrel{i,j,k=1,2,3}{ ({\rm mod} 3\a)}}
\e^{ijk} \frac{\del g_1}{\del q^i(\s)}  
\frac{\del g_2}{\del q^j(\s)} \frac{\del g_3}{\del q^k(\s)}.
\ee
As remarked in the introduction, the 3-bracket \eq{Nfff} 
does not satisfy the fundamental identity.
However due to its simple form and also because it is the most natural 
field theoretic generalization of the canonical Nambu bracket, 
we will refer to it as a Nambu
bracket. In this paper, 
we will not insist on the fundamental identity as a property of
Nambu bracket.

Now we come back to the derivation of \eq{HJ1}. We  note that 
the equation of motion  is equivalent to
\be
F_{ijk} \dot{q}^k - \del_i {\cal H}_1^{(a)} \del_j {\cal H}_2^{(a)} 
+ \del_j {\cal H}_1^{(a)} \del_i {\cal H}_2^{(a)}
= \eps_{ijk} A q^k{}' 
\label{EOM2a}
\ee
for some arbitrary function $A$. 
The presence of an undetermined function $A$
means that the time evolution of $q^i$ 
is not well-defined before gauge fixing. 
This is clear since the EOM \eq{EOM1a} is invariant  
under an arbitrary variation of $\dot{q}$ of the form
\be \label{q-symm}
\dot{q}^i \to \dot{q}^i + A q^i{}' 
\ee
where $A$ is an arbitrary function of $q^i$ and their 
derivatives. This arbitrariness is a direct reflection of 
the gauge symmetry \eq{diffeoa1}. 
Another consequence of \eq{diffeoa1} is that 
the Poisson bracket cannot be defined 
unless the gauge symmetry \eq{diffeoa1} is fixed,
say, by 
$q^1 = \sigma$.

To derive \eq{HJ1},
Takhtajan considered the coordinate transformation (\ref{diffeoa1}) 
with 
\be \label{coord-choice}
\s' = \s + B(t, \s), 
\ee
and 
chose $B$ to satisfy 
\be
\frac{\dot{B}}{1+B'} = \frac{1}{2}f^{-1}_a A.
\ee
Then 
\be
\frac{\del}{\del \s'} = \frac{\del}{\del \s}, \qquad 
\frac{\del}{\del t'} =
\frac{\del}{\del t} - \frac{1}{2} f^{-1}_a A \frac{\del}{\del \s}. 
\ee
As a result, (\ref{EOM2a}) becomes 
\be
\label{gHJ}
\frac{\del q^i}{\del t'} - \{H_1^{(a)},  H_2^{(a)}, q^i \}= 0, 
\ee
which is precisely the 
Nambu-Hamilton equation 
(\ref{HJ1}) in the new coordinate system $(t', \s')$. 
We note that the choice of coordinates \eq{coord-choice} 
is a partial gauge fixing of the 
diffeomorphism symmetry \eq{diffeoa1}. 
The residual gauge symmetry is 
\be
t \rightarrow t, \qquad \sigma \rightarrow \sigma + \d\sigma(\sigma). 
\label{residual_gauge_symm}
\ee
We also note that  if we do not carry out the partial gauge fixing as above, 
the equations of motion (\ref{EOM2a}) can be written as
\be
D q^i(\s) - \{ H_1^{(a)}, H_2^{(a)}, q^i(\s) \} = 0, 
\ee
where 
\be
D \equiv \frac{\del}{\del t} - A \frac{\del}{\del \s}
\ee
resembles the form of a covariant derivative, 
except that $A$ is not a fixed function nor a dynamical variable. 
Finally we note that it is straightforward to generalize the above 
analysis to higher order brackets.


Summarizing, we have shown that when the diffeomorphism symmetry of the 
Takhtajan action \eq{S1a} is partially gauge fixed,
the dynamical evolution of the
system is determined by the  Nambu-Hamilton equation \eq{gHJ}. In this description,
unlike the usual canonical formulation using Poisson bracket, 
the Nambu bracket \eq{Nfff} appears. Even through the bracket \eq{Nfff} does not
satisfy the fundamental identity, the fact that it appears so naturally 
and universally in a class of two dimensional theories suggests that it is 
the right
generalization of the finite dimensional canonical Nambu bracket
\eq{cN}  to the field theoretic
setting.

\subsection{Symmetry algebra of the  Takhtajan action}

\label{symm}

\subsubsection{The case of $D=3$}

The symmetry  of the  
Takhtajan action \eq{S1a} depends 
crucially on the form of $\om$ (\ref{om-def}). 
Since $f_{ijk}$ takes the block-diagonal 
canonical form, let us first discuss the case of $D=3$.

\begin{enumerate}

\item
For $\om = 0$, 
the symmetry algebra of the system is given by the full 
volume preserving diffeomorphism (VPD).

\item
For $\om \neq 0$, 
the symmetry group is an infinite dimensional Abelian subgroup of the VPD.


\end{enumerate}

Without loss of generality,
one can assume that the 3-bracket is given by (\ref{cN-field})
\be
\{ q^i(\s), q^j(\s'), q^k(\s'') \} = \eps^{ijk}\delta(\s-\s')\delta(\s'-\s'')
\ee
by scaling $f_1$ to $1$.
For this case, the Nambu bracket (\ref{cN}), 
which satisfies the fundamental identity, 
and will be denoted as
\be
( q^i, q^j, q^k ) = \eps^{ijk}
\label{cN-new}
\ee
here to distinguish it from the 3-bracket defined above,
has some advantage over (\ref{cN-field}).
When we consider functions of $q_i$ but not their derivatives,
there is a simple connection between the two types of 3-brackets
\be
( {\cal F}_1, {\cal F}_2, {\cal G}_1 ) = {\cal G}_2 
\qquad
\leftrightarrow 
\qquad
\{ F_1, F_2 , {\cal G}_1 \} = {\cal G}_2,
\ee
where $F_i = \int d\sigma \, {\cal F}_i$.

\noindent \underline{Symmetry algebra for $\om = 0$}

Using the fundamental identity 
for (\ref{cN-new}), 
it is easy to see that 
the action (\ref{S1a}) is invariant under 
the  transformations
\be \label{tr}
L({\cal F}_1, {\cal F}_2) \equiv ({\cal F}_1, {\cal F}_2, \ast ) 
= \{ F_1, F_2, \ast \}, 
\ee
where the $F_i$'s are 
integrals of arbitrary functions 
${\cal F}_i$ 
of $q^i \, (i = 1, 2, 3)$.
The symmetry algebra generated by the 
transformations \eq{tr} is indeed the same as the algebra of VPD. To
see this, we note that
for a given pair (or pairs) of functions $({\cal F}_1, {\cal F}_2)$, 
we can define a 1-form in the target space 
${\cal A} = {\cal F}_1 d {\cal F}_2$.
The 
transformation $L$ depends only on 
the ``field strength'' ${\cal F} = d{\cal A}$.
Denoting the 1-form dual to ${\cal F}$ in the target space by $\phi({\cal F})$, 
\be
\phi^i({\cal F}) = \frac{1}{2} \eps^{ijk} {\cal F}_{jk},
\ee
we find $\phi$ to be divergenceless $\del_i \phi^i = 0$,
and thus we can identify $\phi$ with 
the transformation parameter for a volume-preserving diffeomorphism
\be
\d q^i = \phi^i.
\label{VPDdx}
\ee
The duality between ${\cal F}$ and $\phi$ induces 
a 1-1 correspondence between $\phi$ and $L$
\be
L(\phi) = \phi^i \del_i,
\ee
and thus one can easily check that the algebra 
\be
[L(\phi_1), L(\phi_2)] = L(\phi_3),
\ee
where $[\ast,\ast]$ denotes the commutator,
agrees with the algebra of VPD
\be
\phi_3^i = \phi_1^j \del_j \phi_2^i - \phi_2^j \del_j \phi_1^i.
\label{VPD-alg}
\ee

\noindent \underline{Symmetry algebra for $\om \neq 0$}

It is easy to see that the action \eq{S1a}  
is still invariant under the transformation 
\eq{tr} if ${\cal F}_1, {\cal F}_2$ are  constants of motion.
The constants of motion can be constructed easily.  First,
from the 
Nambu-Hamilton 
equation (\ref{HJ1}), a constant of motion
${\cal C}$ has to satisfy the  equation 
\be \label{VC}
{\cal V}^i \del_i {\cal C} =0,
\quad
\mbox{where} 
\quad 
{\cal V}^i \equiv \e^{ijk} \del_j {\cal H}_1^{(a)} \del_k {\cal H}_2^{(a)}.
\ee
This means that $\del_i {\cal C}$ is perpendicular to the 3-dimensional 
vector ${\cal V}^i$.
Therefore, in general one has two independent constants of motion, 
denoted as, say, ${\cal C}_1$ and ${\cal C}_2$.
For example, for the case of $N=1$, one can take 
${\cal H}_1$ and ${\cal H}_2$  as the independent constants of motion.  
It follows that the symmetry transformation of the action can be written as 
\be \label{LF1F2}
L({\cal F}_1, {\cal F}_2) =\frac{\del ({\cal F}_1,{\cal F}_2)}
{\del({\cal C}_1,{\cal C}_2)} 
( {\cal C}_1,{\cal C}_2, \ast ).
\ee
This takes a functional derivative 
in the direction perpendicular to both $d {\cal C}_1$ and $d {\cal C}_2$
in the target space.
Thus all the transformations $L({\cal F}_1, {\cal F}_2)$ 
commute with each other 
and the symmetry algebra is Abelian.

\if
First we have 
the Noether charge for the time translation symmetry 
\be
\label{Ham}
\int d\s \, H_1 H'_2. 
\ee
This is also the Hamiltonian for the system 
in the canonical formulation (for any gauge fixing condition).
In addition there are infinitely many other Noether charges corresponding to 
the diffeomorphism \eq{residual_gauge_symm} of $\s$.
In fact from the 
Nambu-Hamilton equation (\ref{HJ1}), 
it is obvious that $H_1$ and $H_2$ 
are both constants of motion.
This leads to an infinite number of diffeomorphism-invariant conserved quantities 
\be
{\cal O}(F_1, F_2) = 
\int F_1(H_1, H_2 ) dF_2(H_1, H_2),
\ee
where $F_1, F_2$ are arbitrary functions.
 
We note that the corresponding transformation operators
\be 
L(F_1, F_2) = \{ {F}_1, { F}_2, \ast \},
\label{LF1F2}
\ee
takes a functional derivative 
in the direction perpendicular to both $d H_1$ and $d H_2$
in the target space.
Thus all the transformation operators $L(F_1, F_2)$ 
commute with each other 
and the symmetry algebra is Abelian.
\fi

\subsubsection{Higher dimension $D >3$} 
 
As we discussed above, the fundamental identity is no longer valid   
for $D>3$. Nevertheless, the action is invariant under a smaller class of 
special transformations
\be
L({\cal F}_1, {\cal F}_2) \equiv ( \cF_1, \cF_2, \ast ),
\ee
where ${\cal F}_1, {\cal F}_2$ depend only on a 3 dimensional subset of coordinates
$q_i$, $ i\in (1+ 3 \a, 2+ 3 \a, 3+ 3\a)$ 
in which the 3-bracket \eq{Nfff} is block-diagonalized. 
As a result, the symmetry algebra is given by a direct 
sum of the symmetry algebras:
\be
\cA = \oplus_\a \cA_\a,
\ee
where, in the case of $\om =0$, $\cA_\a$'s are 
the VPD of the 3-manifold with coordinates  
$q_i$, $ i\in (1+ 3 \a, 2+ 3 \a, 3+ 3\a)$; and in the case of $\om \neq 0$, 
$\cA_\a$'s are the $U(1)$ symmetry \eq{LF1F2} generated by constants of motion
${\cal C}_1, {\cal C}_2$ defined by \eq{VC} with 
$i,j,k \in (1+ 3 \a, 2+ 3 \a, 3+ 3\a)$.


\if 
As a simple example, 
consider the action
\be
S = \int \left( z dx dy - H_1 d H_2 dt \right)..
\ee
The symmetry generators (\ref{LF1F2}) are 
\be
L(F_1, F_2) = \int d\s \, \{F_1, F_2\} \frac{\d}{\d z(\s)},
\ee
where $F_1, F_2$ are functions of $H_1, H_2$ and $\sigma$ (but not $z$), 
and the bracket is defined by
\be
\{F_1, F_2\} \equiv \frac{\del F_1}{\del x}\frac{\del F_2}{\del y}
- \frac{\del F_1}{\del y}\frac{\del F_2}{\del x}.
\ee
Apparently the algebra of $L(F_1, F_2)$ is Abelian, 
and they generate the following transformations
\be
\d x = 0, \qquad \d y = 0, \qquad 
\d z = \xi(x, y, \s), 
\ee
for an arbitrary function $\xi$. 

The generalized Hamilton-Jacobi equations are 
\be
\dot{x} = \dot{y} = 0, \qquad 
\dot{z} = 1
\ee
for this system, 
and the general solutions are 
\be
x = x_0(\s), \qquad y = y_0(\s), \qquad 
z = t + z_0(\s), 
\ee
for arbitrary functions of $\s$: $x_0(\s), y_0(\s), z_0(\s)$.
For this example, 
we see that the knowledge of $H_i$ being constants of motion, 
and that $L(F_1, F_2)$ generates a symmetry of the system, 
constitute all the information about the space of solutions. 

Some of the solutions are equivalent due to 
the residual gauge symmetry (\ref{residual_gauge_symm}).
We can fix gauge by setting one of 
the three functions $(x_0, y_0, z_0)$ 
to a given function of $\s$, say, $z_0 = z - t =\s$.
\fi

\end{document}